\documentclass[letterpaper,twocolumn,10pt]{article}
\usepackage[hyphens]{url}
\usepackage{soul}
\usepackage{usenix,epsfig,hyperref,caption,subcaption}
\hypersetup{
    bookmarksnumbered=true,
    unicode=false,          % non-Latin characters in AcrobatÕs bookmarks
    pdftoolbar=true,        % show AcrobatÕs toolbar?
    pdfmenubar=true,        % show AcrobatÕs menu?
    pdffitwindow=false,     % window fit to page when opened
    pdfstartview={FitH},    % fits the width of the page to the window
    pdftitle={A $35 Firewall for the Developing World},    % title
    pdfnewwindow=true,      % links in new window
    colorlinks=false,       % false: boxed links; true: colored links
    linkcolor=red,          % color of internal links
    citecolor=green,        % color of links to bibliography
    filecolor=magenta,      % color of file links
    urlcolor=cyan           % color of external links
}

\begin{document}
%
% --- Author Metadata here ---
%\conferenceinfo{NSDR}{'14 El Paso, Texas USA}
%\CopyrightYear{2007} % Allows default copyright year (20XX) to be over-ridden - IF NEED BE.
%\crdata{0-12345-67-8/90/01}  % Allows default copyright data (0-89791-88-6/97/05) to be over-ridden - IF NEED BE.
% --- End of Author Metadata ---

\title{\Large \bf A \textbf{\$}35 Firewall for the Developing World}

\author{
{\rm Zubair Nabi}\\
IBM Research, Dublin\\
zubairn@ie.ibm.com
%Paper \# 3
}

\maketitle

\thispagestyle{empty}

\begin{abstract}
A number of recent efforts aim to bridge the global digital divide, particularly
with respect to Internet access. We take this endeavor one step further and
argue that Internet access and web security go hand in glove in the developing
world. To remedy the situation, we explore whether low-cost platforms, such as
Raspberry Pi (\$35) and Cubieboard (\$59), can be used to implement security
mechanisms. Using a firewall as a motivating security application we benchmark
its performance on these platforms to test our thesis. Our results show that
these platforms can indeed serve as enablers of security functions for small
sized deployments in the developing world, while only consuming less than \$2.5
worth of electricity per device per annum. In addition, we argue that the use of
these platforms also addresses maintenance challenges such as update roll-out
and distribution. Furthermore, a number of additional network functions, such as
caching and WAN acceleration can also be implemented atop this simple
infrastructure. Finally, we posit that this deployment can be used for
in-network monitoring to facilitate ICT4D research.
\end{abstract}

%\category{C.2.0}{Computer-Communication Networks}{General}[Security and
% protection]
%A category including the fourth, optional field follows...
%\category{C.2.3}{Network Operations}{Network management}

%\terms{Design, Human Factors, Performance}

%\keywords{Developing World, Middleboxes, Security, ICT4D}

\section{Introduction}
Over the course of the last few decades, the Internet has matured into a
repository of human knowledge, a medium for dissemination of ideas, and more
generally, an all-encompassing portal for planet-scale connectivity. It has also
become an integral part of the global economy. So much so that in the period
between 2006 and 2011, it accounted for 21\% of GDP growth in developed
countries~\cite{Manyika:2011:TGT}. In addition, maturity in the Internet
ecosystem has resulted in a higher standard of life~\cite{Manyika:2011:TGT}. In
the same vein, Internet access coupled with social media has become a catalyst
for social, cultural, and political activism and
change~\cite{Starbird:2012:RRI,Vieweg:2010:MDT,Sullivan:2009:EAS,Yang:2013:Power}.
While the Internet has been declared a basic human
right~\cite{Sathiaseelan:2013:LLC}, in reality more than two-thirds of the world
population---which lives on less than two dollars a
day~\cite{Foo:2010:CCS}---does not have access to it. This Internet blackout can
be attributed to a set of technological, social, and economic factors.

To bridge this connectivity gap, researchers, social entrepreneurs, and industry
specialists have explored and deployed a number of radical solutions.
These range from wireless networks driven by
WiMAX~\cite{Lele:2007:PVC,Chaudhari:2012:MGB},
satellite~\cite{Matthee:2007:BIC}, ZigBee~\cite{Raman:2009:Lo3}, long-distance
WiFi~\cite{Patra:2007:WDI}, wireless mesh~\cite{Johnson:2007:EOA}, and cellular
links~\cite{Heimerl:2013:ERC} to wired technologies enabled by optical, dial-up,
and analog cable networks~\cite{Nabi:2013:RLC}. These backbone and last mile
access network technologies are augmented by a similarly rich array of
conventional and unconventional optimizations, including aggressive
caching~\cite{Badam:2009:HCS}, prefetching and offline
access~\cite{Pai:2010:FAD}, P2P content sharing~\cite{Saif:2007:PMB}, and
village-level kiosks~\cite{Brewer:2006:CTR,Lele:2007:PVC}.
Unfortunately, Internet backbone connectivity is still a bottleneck factor due
to its high-cost~\cite{Saif:2007:PMB}. This coupled with the limited data rate
of some of these technologies, restricts end-user connectivity to the kilobit
order. Cognizant of the potential of such a large untapped market, technology
giants have also recently jumped into the fray, with Google and Facebook leading
the way with Project Loon~\cite{Google:Loon} and
Internet.org~\cite{Zuckberberg:ICA}, respectively.

% The challenges in providing connectivity in the developing world are further
% exacerbated by obstacles unique to the local environment and conditions.
% Specifically, electricity coverage is limited~\cite{Foo:2010:CCS} and where
% available, widespread power cuts are the norm. To make matters worse, the power
% experiences high fluctuation leading to frequent equipment
% failure~\cite{Brewer:2005:CTD}. This enforces a high overhead of regular
% component replacement and back up power sources. Further, extreme environmental
% conditions such as heat and dust aggravate the situation. Moreover, network
% layout and design is non-trivial due to the low population
% density~\cite{Blantz:2012:THR}. Finally, maintenance of the infrastructure needs
% to be automated due to the lack of a local skilled workforce in these
% regions~\cite{Brewer:2006:CTR}.

Above all, the network security model in the developing world is considerably
different than what the security research community and the technology industry
has hitherto focused
on~\cite{Ben-David:2011:CSD,Bhattacharya:2011:CVU,Paik:2011:GCE}. This is
evinced by the disproportionally high rate of cybercrime originating in
developing countries. In addition, these countries reside on the higher end of
the global spam scale as well as botnet activity~\cite{Ben-David:2011:CSD}. On
the one hand, these problems weaken the security and resilience of the worldwide
Internet infrastructure and on the other, they hamper widespread deployment by
denting the confidence of the average user in technology. According to Ben-David
\emph{et al.}, the multi-faceted factors specific to developing countries
include lack of regular online software and firmware updates due to limited
bandwidth, shared computing resources, low-literacy of the users, and rampant
software piracy~\cite{Ben-David:2011:CSD}.

\subsection{Another Brick in the Firewall}

Viruses are especially uncontrollable in Internet cafes---which are a primary
source of connectivity for most users in the developing world---due to shared
USB flash drives, untrained users, and limited financial and human
resources~\cite{Bhattacharya:2011:CVU}. Some researchers have gone to the extent
of arguing that virus ecology and epidemiology in the developing world is
fundamentally different than the developed world~\cite{Paik:2011:GCE}.
Furthermore, the networks in these regions are largely insecure due to the high
cost of enterprise-grade middleboxes such as firewalls and thus the networks are
susceptible to even simple port scans. Fortunately, the research community has
started pushing for generalized middleboxes~\cite{Sekar:2011:MME}, although the
target so far has been high-end applications~\cite{Sherry:2012:MMS}.

In this paper, we explore how the recent calls for middlebox innovation can be
leveraged to break the security status quo in the developing world.
Specifically, we try to ascertain whether low-cost platforms such as the
Raspberry Pi~\cite{Halfacree:2012:RPi} and Cubieboard~\cite{Cubieboard} can be
used as middleboxes to implement firewall functionality to protect alternative
network deployments or small Internet cafe level LANs. These networks include
those supported by long-distance WiFi, WiMAX, and Zigbee, to name a few. While
we benchmark the performance of a firewall application on these two platforms,
our thesis is in no way limited to them. The case is equally applicable to other
similar platforms such as Utilite~\cite{Utilite}, Arduino~\cite{Arduino}, and
BeagleBoard~\cite{Beagle}. In fact, as we discuss further on, platforms such as
the NetFPGA are also viable options. Furthermore, the platforms can also be used
to provide other security services such as local software upgrade patches and
intrusion detection systems as well as more general middlebox applications such
as content caching and traffic shaping.

The rest of the paper is organized as follows. In \S\ref{sec:background} we
give an introduction to our target alternative networks and low-cost platforms.
\S\ref{sec:application} presents our target application and its evaluation on
two low-cost platforms. General use-cases and platforms are discussed in
\S\ref{sec:discussion}. We summarize relevant related work in
\S\ref{sec:related} and finally conclude in \S\ref{sec:conclusion}.

\section{Background}\label{sec:background}
In this section, we first present alternative networks which have been designed
specifically for the developing world and then analyse various low-cost,
single-chip devices.

\begin{table}[t]
\centering
    \begin{tabular}{| l | l |}
    \hline
    \textbf{Technology} & \textbf{Net Bandwidth (Mbps)}\\
	\hline
    ZigBee~\cite{Raman:2009:Lo3} & 0.060\\
    Satellite~\cite{Matthee:2007:BIC} & 1\\
    Wireless mesh~\cite{Johnson:2007:EOA} & 2.5\\
    Long-distance WiFi~\cite{Patra:2007:WDI} & 5\\
    WiMAX~\cite{Chaudhari:2012:MGB} & 6\\
    \hline
    \end{tabular}
    \caption{Comparison of Alternative Networks}
    \label{tab:com}
\end{table}

\subsection{Alternative Networks}
Alternative networks augment existing technologies by customizing them to
support low-cost, low-power, and low-maintenance. Table~\ref{tab:com} lists the
solutions that have been deployed in various locations around the world and
their data rates. We discuss these in detail in this section.

\paragraph{Long-distance WiFi}
Long-distance WiFi initiatives extend the range of the specification by
modifying the MAC layer. One such implementation, dubbed
\emph{WiLDNet}~\cite{Patra:2007:WDI}, addresses three shortcomings in the
vanilla 802.11 protocol for long-distance communication: 1) sub-optimal
link-level recovery, 2) frequent collisions due to CSMA/CA, and 3) inter-link
interference. To this end, it uses bulk packet acknowledgement, TDMA enabled by
loose time synchronization, and adaptive loss-recovery.

\paragraph{ZigBee}
Lo3, which stands for ``Low-cost, Low-power, Local communication'', advocates
the use of 802.15.4 for rural connectivity~\cite{Raman:2009:Lo3}. The use of
802.15.4 enables the setup to minimize its energy footprint by consuming power
on the $\mu$W and mW scale during idle and normal operation, respectively. To
negate investment in a centralized tower, Lo3 makes use of a mesh network in
which the medium is arbitrated by centralized TMDA.

\paragraph{Satellite}
Satellite networks have also been employed for backbone connectivity in rural
areas. For instance, in rural Zambia, VSAT (Very Small Aperture Terminal)
satellite connections are being used to provide Internet
connectivity~\cite{Matthee:2007:BIC}. This bandwidth is then distributed through
a three-tier WLAN within the community: one main tower (wide-area backbone)
connected to the VSATs and peered with other towers (local-area backbone), which
in turn provide connectivity to end-hosts through Ethernet and wireless access
points.

\paragraph{WiMAX}
WiMAX greatly reduces the cost of network deployment and also increases its
reach to rural areas where the geographic terrain is not amenable to copper and
optical wiring. Lele \emph{et al.}~\cite{Lele:2007:PVC} advocate the use of such
a deployment in rural India, where a single WiMAX base station serves the entire
community. Architecturally, it revolves around a kiosk model in which end-users
with regular telephone sets are connected to kiosks which in turn communicate
with the base station.

\paragraph{Wireless Mesh}
As opposed to mainstream urban networks---which depend on well planned antenna
configurations and nodes with multiple radios---networks designed for rural
areas in the developing world rely on nodes with single
radios~\cite{Johnson:2007:EOA}. This ensures low cost and simplicity which are
first-class goals for these networks. The single radio nodes are connected in
the form of a mesh to provide connectivity at the community level. End-hosts
therefore are connected through multiple hops to the gateway.

\subsection{Basic Platforms}
In this section, we first give a feature-set overview of the Raspberry
Pi~\cite{Halfacree:2012:RPi} followed by Cubieboard~\cite{Cubieboard}.

\begin{table}[t]
\centering
    \begin{tabular}{| l | l | l |}
    \hline
    \textbf{Component} & \textbf{Raspberry Pi} & \textbf{Cubieboard}\\
     & \textbf{(Model B)} & \\
	\hline
    Processor (MHz) & 700 & 1000\\
    GPU (GLFOPS) & 24 & 7.2\\
    Memory (MB) & 512 & 1024\\
    Ethernet & 10/100 & 10/100\\
    \hline
    Price (\$) & 35 & 59\\ 
    \hline
    \end{tabular}
    \caption{Feature-based Comparison of Platforms}
    \label{tab:plat:com}
\end{table}

\paragraph{Raspberry Pi}
Raspberry Pi is powered by a Broadcom system-on-chip multimedia processor with
an ARM 1176JZF-S 700MHz processor and a VideoCore IV 24GFLOPS GPU. For storage
it relies on an external MMC or SD Card. Available in two models (A:
256MB RAM, 1 USB port, and B: 512MB RAM, 2 USB ports, and 10/100 Ethernet), it
can be interfaced with external components via GPIO (General-purpose
Input/Output) and UART (Universal Asynchronous Receiver/Transmitter). On the
software side, a number of popular Linux distributions, such as Debian and
Fedora have Raspberry Pi specific versions.

\paragraph{Cubieboard}
Cubieboard consists of an AllWinner A10 system-on-chip processor with an ARM
Cortex-A8 1GHz processor and a Mali 400 7.2GFLOPS GPU. For storage, it relies on
both built-in 4GB NAND Flash and external slots for microSD and SATA.
Moreover, it has 1GB RAM and 10/100 Ethernet. External interfacing is enabled by
2X USB slots, $I^{2}C$ (Inter-Integrated Circuit), SPI (Serial Peripheral
Interface), and LVDS (Low-voltage Differential Signaling). Similar to the
Raspberry Pi, the Cubieboard is also driven by Debian and Fedora based
distributions.

Table~\ref{tab:plat:com} compares the key features of both platforms.

\begin{table}[t]
\centering
    \begin{tabular}{| l | l |}
    \hline
    \textbf{Chain} & \textbf{Function}\\
	\hline
    \texttt{PREROUTING} & Pre-routing decision packets\\
    \texttt{POSTROUTING} & Post-routing decision packets\\
    \texttt{INPUT} & Incoming packets for local delivery\\
    \texttt{OUTPUT} & Outgoing packets\\
    \texttt{FORWARD} & Incoming packets for non-local delivery\\ 
    \hline
    \end{tabular}
    \caption{Default \texttt{iptables} chains}
    \label{tab:chains}
\end{table}

\section{Motivating Application}\label{sec:application}
Developing world countries---such as India, China, and Brazil---are top
sources of spam and botnet activity. To make matters worse, the number of
botnets in these countries is expected to exceed the developed world soon due to
the increase in digital connectivity as well as the poor security hygiene of the
area~\cite{Klaar:2013:bcn}. Therefore, in the future, Internet security
battle-lines will be drawn in the developing world.

Traditionally, firewalls have been employed to stem the botnet tide. Firewalls
provide a first line of defense on the network against malicious activity. Each
firewall has a list of rules that it enforces to keep unwanted traffic at bay.
These rules decide if a certain flow is to be accepted or rejected. In addition,
these rules can work at multiple levels in the protocol stack from the link
layer up to the application layer. Naturally, the efficacy of any firewall is
determined by its ruleset. Multiple rules in tandem can be used (in the form of
a rule chain) to implement complex filtering policies. A key requirement is the
ability to filter packets at line-rate otherwise the network experiences a
decrease in QoS. Therefore, enterprise-grade firewalls are designed to achieve
fast processing while providing rich features such as policy-based filtering and
deep packet inspection. Unfortunately, enterprise-grade firewalls (and
middleboxes in general) are cost-prohibitive. At the other end of the spectrum,
general-purpose machines can be employed to perform basic firewalling.

The aim of this paper is to explore the use of the most basic platforms for
firewall functionality. To this end, we use real-world firewall rule set
cardinality to gauge the throughput of low-cost platforms to protect alternative
networks in the developing world. These platforms can be considered as an
example of commodity off-the-shelf hardware which is more appropriate for these
regions~\cite{Patra:2007:WDI}. Architecturally, we envision these firewalls to
be present at the backbone of alternative networks such as those supported by
long-distance WiFi or at the network gateway of Internet cafes. According to
figures from 2009, the average firewall contains 800 rules and the maximum
number in the wild is 20000~\cite{Chapple:2009:AAO}. Therefore, we use these
numbers for our benchmarking.

\begin{figure}[t]
\centering
  \includegraphics[width=1\linewidth]{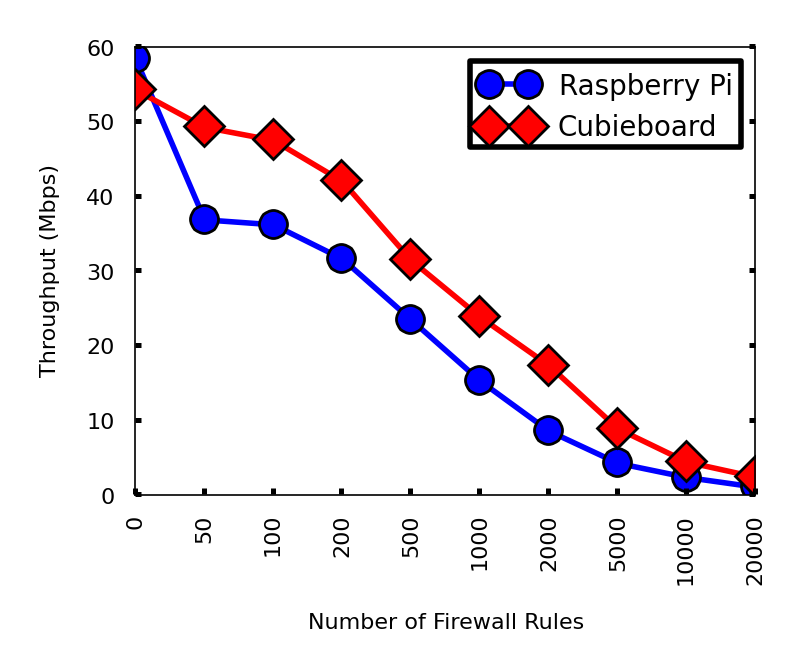}
  \vspace{-10pt}
  \caption{Raspberry Pi vs. Cubieboard}
  \label{fig:results}
  \vspace{-10pt}
\end{figure}

\subsection{Setup and Results}
In this section, we present the results from our evaluation of Raspberry Pi and
Cubieboard. We used the standard \texttt{iptables} application in Linux as our
test firewall. \texttt{iptables} allows the user-space configuration of
\emph{chains} of rules (default chains enumerated in Table~\ref{tab:chains}) of
the kernel firewall (\texttt{Netfilter}) for IPv4. Stateful rules that
\texttt{ACCEPT}, \texttt{REJECT}, \texttt{DROP}\footnote{As opposed to
\texttt{REJECT} in which the source is notified, \texttt{DROP} silently drops
the packet without any source notification.}, or \texttt{LOG} a packet can be
added to these chains.

By default, the \texttt{iptables} module is not available in the stock kernel
disk image of both platforms. Therefore, we linked the module in and recompiled
the kernel for a Linux distribution for each platform. For Raspberry Pi we used
a Wheezy Raspbian image with kernel version 3.6.11 while for Cubieboard we used
a disk image from Linaro with kernel version 3.10.1. As our goal was to measure
the performance in terms of flow forwarding of each platform subject to firewall
filtering, we used \texttt{iperf}\footnote{\texttt{iperf} allows the measurment
of throughput for TCP and UDP streams between two end-hosts.} to measure
throughput between the platform and a standard host (a single hop away). We used
the default settings for \texttt{iperf} which use TCP as the transport with a
window size of 16KB. A custom script was used to add random rules to
\texttt{iptables}. Figure~\ref{fig:results} plots the throughput of each
platform as a function of the number of firewall rules. Without any rules,
Raspberry Pi has a throughput of around 58Mbps as opposed to 54Mbps for
Cubieboard. As we increase the number of rules, the performance of Cubieboard
degrades more gracefully in comparison to Raspberry Pi. This is due to the
higher processing power of Cubieboard (1GHz) versus Raspberry Pi (700MHz). While
relative numbers are immaterial for our cause, the analysis shows that both
platforms are capable of sufficient throughput (20Mbps Raspberry Pi and 30Mbps
Cubieboard) for an average number of firewall rules (800) in the
wild~\cite{Chapple:2009:AAO}.

\section{Discussion}\label{sec:discussion}
The use of low-cost platforms simplifies update rollout and also enables
additional functions such as content caching and remote monitoring while
ensuring a low energy footprint. We discuss these and other topics in this
section.

\subsection{Update Rollout}
As mentioned earlier, one of the challenges in providing security (or any
technology) in the developing world is the low-literacy of both service
providers as well as end-users. In addition, the high cost of maintenance is a
major impediment. Therefore, it is useful to have systems that can be easily be
upgraded and maintained. To this end, SD Card support on low-cost platforms
simplifies application and ruleset rollout. For instance, the OS pre-loaded with
firewall rules can be supplied as an SD Card image and quickly disseminated
through self-replication~\cite{Monteil:2013:SHS}. In addition, extra rules can
be downloaded from the Internet on the fly as rules specific to the local
network evolve.

\subsection{Higher Performance}
For alternative networks, the 30Mbps for an average number of firewall rules
suffices. For networks with higher bandwidth requirements, NetFPGA-like
platforms can be employed. NetFPGA-1G consists of 4 x 1Gb interfaces while the
recent 10G version features 4 x 10Gb ports. The customized hardware (PCI board
with FPGA) and firewall software can enable near line-rate processing.

\subsection{Additional Functions}
The extra CPU cycles and storage on the platform can be employed for additional
applications. For instance, the storage can be used to provide network wide
caching~\cite{Badam:2009:HCS}. This will assist in keeping bandwidth usage in
check. Moreover, due to the volatility of these networks, connectivity is
intermittent. In such situations, popular content can be served from the
firewall platform for temporary offline access. Other potential applications
include WAN acceleration and pre-fetching~\cite{Pai:2010:FAD}.

\subsection{Distributed Firewalls}
Rule management can be further simplified by enabling distributed
firewalls~\cite{Ioannidis:2000:IDF}. In this architecture, we envision a
centralized service which maintains lists of firewall policies. These lists can
be distributed based on geography or network behaviour similarity.
These rules can then be rolled out to Raspberry Pi or Cubieboard-enabled
platforms via IPSec mechanisms, for enforcement at the edge.

\subsection{NAT}
An additional function that firewalls afford is network address translation.
NATs assist in obfuscating internal addresses. Internet cafes which are still
thriving in the developing world can benefit from this in two key ways: 1) they
can protect internal machines from malicious traffic and 2) they can enable
simple subnet allocations for devices behind the NAT.

\subsection{Monitoring}
We envision that each local device will also be connected to a central
server\footnote{We do not necessarily mean a single server per se but
potentially a geo-distributed architecture that maintains either local or global
network state.} and will periodically communicate different network statistics.
This monitoring will help in understanding the dynamics of firewalls in the
developing world and in devising new firewall rules. In addition, these
statistics will also enable ICT4D practitioners and researchers to understand
the usage patterns of devices and users in the wild. Finally, remote monitoring
will aid in diagnosing network-level problems which are beyond the skill-set of
the local workforce~\cite{Brewer:2006:CTR}.

\subsection{Energy Footprint}
As power is a constrained resource in the developing
world~\cite{Brewer:2006:CTR}, it is imperative to minimize the energy footprint
of any additional infrastructure. Fortunately, Raspberry Pi-like platforms draw
minimal power. Raspberry Pi Model B has a power rating of 3.5W, therefore it
will draw a maximum of 84W in a day. With an average cost of 8cents/kWh in
India~\cite{Wilson:2013:AEP}, a quick back-of-the-envelope calculation reveals
that the energy cost of an always-on Raspberry Pi Model B powered firewall is
\$0.01 daily or \$2.45 annually, which makes it affordable for widespread
deployment. In fact, it can even be removed from the power grid and powered by
batteries to avoid equipment failure due to high power fluctuation in these
regions~\cite{Brewer:2006:CTR}.

\section{Related Work}\label{sec:related}
Our work is inspired by Ben-David \emph{et al.}~\cite{Ben-David:2011:CSD} who
argue that security issues in the developing world are fundamentally different
than the developed world. In particular, our firewall analysis is motivated by
their insight that this not only has an adverse effect on the global
technological landscape but also hinders technology adoption in the developing
world. In a similar vein,~\cite{Paik:2011:GCE} has also argued for security
solutions designed specifically for the developing world. To this end,
Innoculous is a self-contained tamper-proof anti-virus system on a flash drive
for virus detection and profiling~\cite{Paik:2011:GCE}. Similarly, Bhattacharya
and Thies~\cite{Bhattacharya:2011:CVU} present the dynamics of viruses in
Internet cafes in India. They also highlight a number of research opportunities
including computer virus epidemiology, use of disk imaging for simple roll-back,
taking advantage of loose privacy norms, and leveraging the want of the owners
to pay more for virus reduction, such as subscription based training and updated
virus definitions via post. Our approach of provisioning firewall rules along
with the OS stack as a disk image is a direct consequence of their
recommendations. While all of these proposals target security in the developing
world, to the best of our knowledge this paper is the first attempt at employing
low-cost platforms for network security applications. Finally, our work has
benefited from the rich body of
work~\cite{Lele:2007:PVC,Chaudhari:2012:MGB,Matthee:2007:BIC,Raman:2009:Lo3,Patra:2007:WDI,Johnson:2007:EOA,Heimerl:2013:ERC,Nabi:2013:RLC,Badam:2009:HCS,Pai:2010:FAD,Saif:2007:PMB,Brewer:2006:CTR,Lele:2007:PVC}
that aims at providing low-cost connectivity in the developing world.

\section{Conclusion}\label{sec:conclusion}
We explored the use of low-cost platforms such as Raspberry Pi and Cubieboard
for security applications in the developing world. Our benchmarking of a
firewall proves that the capabilities of these platforms are sufficient to fill
the security void in developing regions. Using an average number of firewall
rules, we showed that a Cubieboard can achieve throughput of 30Mbps. The use of
low-cost platforms also simplifies application rollout in the form of an SD Card
image. In addition, the power draw of the deployment is small enough to make it
feasible and sustainable. We also discussed how the same platform can be used
for general networking applications such as WAN acceleration and content caching
and pre-fetching. Moreover, remote monitoring of these platforms can both aid in
network diagnostics as well as enhancing ICT4D research by laying bare the
characteristics of network traffic and user behaviour in the developing world.

In addition to field-testing and real-world deployment, we are also interested
in writing custom security applications such as firewalls and wrapping each in a
Mirage unikernel~\cite{Madhavapeddy:2013:ULO} and running it directly on
baremetal Raspberry Pi-like platforms. This has two main advantages: 1) It
simplifies application shipment by turning the entire stack into a single image
and 2) It enhances the security of the application itself by making its code
type safe and sealing it against runtime modification.

%\newpage
{\footnotesize \bibliographystyle{acm}
\bibliography{sigproc}}

%\balancecolumns % GM June 2007
% That's all folks!
\end{document}